\documentclass[sigconf]{acmart}

\usepackage{graphicx}

\title{StructInbet: Integrating Explicit Structural Guidance into Inbetween Frame Generation}
\author{Zhenglin Pan}
\affiliation{%
  \institution{Japan Advanced Institute of Science and Technology}
  \country{Japan}
}
\email{z-pan@jaist.ac.jp}

\author{Haoran Xie}
\affiliation{%
  \institution{Japan Advanced Institute of Science and Technology}
  \country{Japan}
}
\email{xie@jaist.ac.jp}

\setcopyright{rightsretained}
\copyrightyear{2025}
\acmYear{2025}
\acmConference{SIGGRAPH Posters '25}{August 10-14, 2025}{Vancouver, BC, Canada}\acmBooktitle{Special Interest Group on Computer Graphics and Interactive Techniques Conference Posters (SIGGRAPH Posters '25), August 10-14, 2025}\acmDOI{10.1145/3721250.3743032}
\acmISBN{979-8-4007-1549-5/2025/08}

\begin{document}



\begin{teaserfigure}
  \centering
  \includegraphics[width=0.99\linewidth]{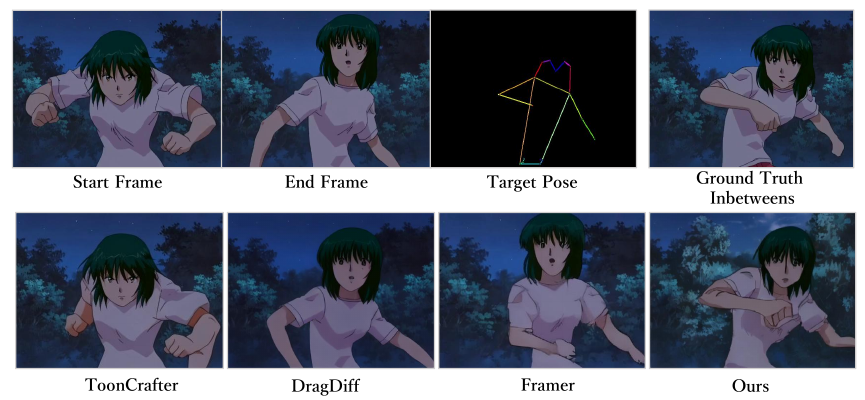}
  \caption{\textbf{Compared to previous interpolation~\cite{xing2024tooncrafter} and drag-based editing methods~\cite{shi2024drag,wang2024framer}, StructInbet enables users to generate inbetween frames with improved pose alignment.}}
  \label{fig:teaser}
\end{teaserfigure}

\maketitle

\section{Introduction}
\label{sec:intro}

Traditional animation studios rely heavily on manual labor to generate inbetween frames between keyframes. Early efforts to automate this task employed optical flow or feature matching techniques, but they offer limited user control. Recent diffusion-based models, guided by interactively dragging points, allow users to control the generation of intermediate frames. Nonetheless, dragging pixels may fail to accurately capture the character gestures, as point-based inputs are inherently sparse and lack structural context.

    To address these issues, we propose \textbf{StructInbet}, an inbetweening system designed to generate controllable transitions over explicit structural guidance. StructInbet introduces two key contributions. First, we propose explicit structural guidance to the inbetweening problem to reduce the ambiguity inherent in pixel trajectories. Second, we adopt a temporal attention mechanism that incorporates visual identity from both the preceding and succeeding keyframes, ensuring consistency in character appearance.

\begin{figure*}[t]
  \centering
  \includegraphics[width=1.0\linewidth]{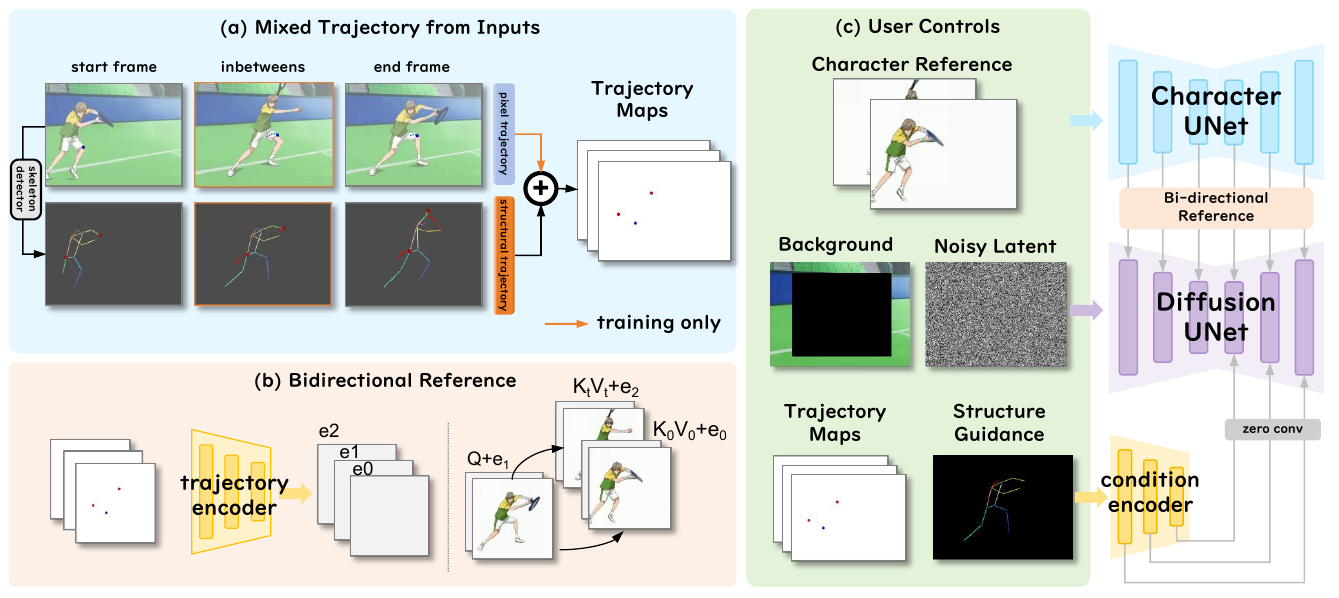}
  \caption{\textbf{StructInbet constructs (a) mixed trajectory maps from input frames and uses structural guidance to condition the diffusion UNet. The Character UNet extracts character features from the surrounding frames and fuses the latent features through (b) bidirectional reference attention. (c) These controls jointly guide the diffusion UNet to synthesize the intermediate frame during inference.}}
  \label{fig:figure1_architecture}
  \vspace{-2ex}
\end{figure*}

\section{Method}
\label{sec:method}

The overall model architecture is shown in Figure~\ref{fig:figure1_architecture}. StructInbet is implemented on a pre-trained Stable Diffusion UNet~\cite{diffusion}, extended with two conditioning branches. The first is a ControlNet~\cite{zhang2023controlnet} encoder that processes structural guidance and trajectory maps, guiding gesture generation in the diffusion UNet through zero-convolution modules. The second is a character reference UNet, architecturally similar to the diffusion UNet, that extracts appearance features from the start and end frames to ensure visual consistency of the character's appearance.

To incorporate diverse guidance signals and enhance temporal coherence, we combine skeleton control with a mixture of pixel trajectories and skeleton joint correspondences to form a set of mixed trajectories that encode both spatial and semantic mappings. While skeletons provide deterministic pose layouts, pixel trajectories enable flexible, fine-grained mapping augmentations. The pixel trajectories are then rasterized into an integer-labeled guidance map, which is concatenated with the skeleton to guide the generation. To maintain appearance consistency over time, we introduce a bidirectional reference attention mechanism. Specifically, a character reference UNet extracts key-value pairs \((K_0, V_0)\) and \((K_T, V_T)\) from the two keyframes and replaces the original keys and values in the self-attention blocks of the diffusion UNet. The attention is calculated as follows:
\begin{equation}
\mathrm{Attn} = \frac{1}{2} \left(
  \mathrm{softmax}\left(\frac{\tilde{Q}\tilde{K}_0^\top}{\sqrt{d}}\right)\tilde{V}_0 +
  \mathrm{softmax}\left(\frac{\tilde{Q}\tilde{K}_T^\top}{\sqrt{d}}\right)\tilde{V}_T
\right)
\end{equation}

We enhance the spatial alignment within the self-attention by incorporating the embeddings of mixed trajectories, produced by a dedicated trajectory encoder. where \(\tilde{Q} = Q + \mathrm{emb}_c\), \(\tilde{K}_0 = K_0 + \mathrm{emb}_0\), \(\tilde{K}_T = K_T + \mathrm{emb}_T\), \(\tilde{V}_0 = V_0 + \mathrm{emb}_0\), and \(\tilde{V}_T = V_T + \mathrm{emb}_T\). Here, \(\mathrm{emb}_c\), \(\mathrm{emb}_0\), and \(\mathrm{emb}_T\) are condition embeddings for the rasterized maps at respective time. This bidirectional attention enables the model to integrate spatial context from two temporal endpoints at a time, improving alignment and character consistency in the generated inbetweens.

\section{Experiments}



\begin{figure}[t]
  \centering
  \includegraphics[width=1.0\linewidth]{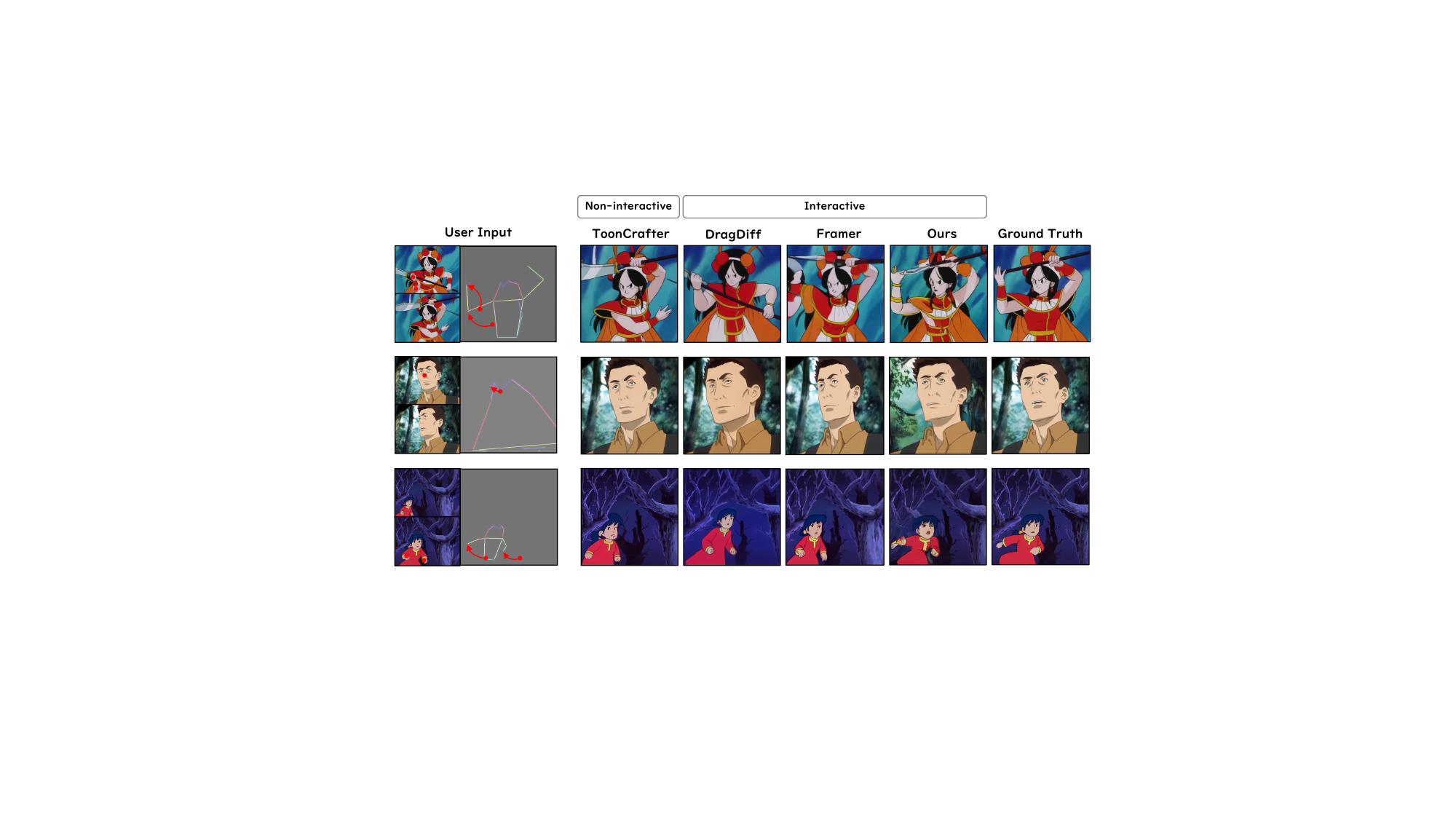}
  \caption{\textbf{Visual comparisons of the produced inbetweening results using state-of-the-art methods and ours.}}
  \label{fig:experiments}
  \vspace{-2ex}
\end{figure}

We compare StructInbet with both non-interactive and interactive baseline methods, as shown in Figure~\ref{fig:experiments}. Non-interactive video interpolation models like ToonCrafter~\cite{xing2024tooncrafter} lack user control, resulting in uncontrollable generation. Among interactive methods, we compare with DragDiffusion~\cite{shi2024drag} and Framer~\cite{wang2024framer}, which support point-based dragging on keyframes but lack explicit pose information needed for detailed or complex pose specification.

Table~\ref{table:overall_comparison} shows both quantitative metrics and user preference results. StructInbet achieves lower FID and LPIPS scores, indicating improved performance under structural guidance. For human evaluation, 6 independent participants rated the results of each method using a 5-point Likert scale across three criteria: image quality(IQ), character consistency(CC), and motion alignment(MA). While ToonCrafter performs well in IQ and CC, it lacks interactive control. Among controllable methods, StructInbet receives significantly higher ratings for structural fidelity while maintaining comparable IQ and CC scores to DragDiffusion. 

\begin{table}[h]
\small
\centering
\caption{\textbf{Quantitative comparison and user preference study.} \\ $\downarrow$: lower is better. $\uparrow$: higher is better.}
\label{table:overall_comparison}
\begin{tabular}{lccccccc}
\toprule
\textbf{Method} & \textbf{FID↓} & \textbf{LPIPS↓} & \textbf{PSNR↑} & \textbf{IQ↑} & \textbf{CC↑} & \textbf{MA↑} \\
\midrule
ToonCrafter    & \textbf{0.29} & 0.56          & 11.06          & \textbf{3.29}  & \textbf{3.57} & 2.14 \\
DragDiffusion  & 1.16          & 0.57          & 11.18          & 2.79           & 2.64         & 2.36 \\
Framer         & 0.45          & 0.56          & 11.24          & 2.79           & 3.07         & 2.71 \\
\textbf{StructInbet} & 0.58 & \textbf{0.54} & \textbf{11.89} & 2.79 & 2.64 & \textbf{3.86} \\
\bottomrule
\end{tabular}
\end{table}

\section{Conclusion}

We presented StructInbet, an inbetweening system designed to generate controllable transitions over explicit structural guidance. Compared with prior implicit point-based approaches, StructInbet enables users to directly manipulate the inbetween generation with greater structural clarity and reduced the overall ambiguity, leading to frame generation that aligns more closely with user intent.

\begin{acks}
This work is funded by New Energy and Industrial Technology Development Organization (NEDO) JPNP20017.
\end{acks}


{\small
\bibliographystyle{ACM-Reference-Format}
\bibliography{ref}


\begin{thebibliography}{5}


\ifx \showCODEN    \undefined \def \showCODEN     #1{\unskip}     \fi
\ifx \showDOI      \undefined \def \showDOI       #1{#1}\fi
\ifx \showISBNx    \undefined \def \showISBNx     #1{\unskip}     \fi
\ifx \showISBNxiii \undefined \def \showISBNxiii  #1{\unskip}     \fi
\ifx \showISSN     \undefined \def \showISSN      #1{\unskip}     \fi
\ifx \showLCCN     \undefined \def \showLCCN      #1{\unskip}     \fi
\ifx \shownote     \undefined \def \shownote      #1{#1}          \fi
\ifx \showarticletitle \undefined \def \showarticletitle #1{#1}   \fi
\ifx \showURL      \undefined \def \showURL       {\relax}        \fi
\providecommand\bibfield[2]{#2}
\providecommand\bibinfo[2]{#2}
\providecommand\natexlab[1]{#1}
\providecommand\showeprint[2][]{arXiv:#2}

\bibitem[Rombach et~al\mbox{.}(2021)]%
        {diffusion}
\bibfield{author}{\bibinfo{person}{Robin Rombach}, \bibinfo{person}{Andreas Blattmann}, \bibinfo{person}{Dominik Lorenz}, \bibinfo{person}{Patrick Esser}, {and} \bibinfo{person}{Bj{\"o}rn Ommer}.} \bibinfo{year}{2021}\natexlab{}.
\newblock \bibinfo{title}{High-resolution image synthesis with latent diffusion models, 2021}.
\newblock
\newblock


\bibitem[Shi et~al\mbox{.}(2024)]%
        {shi2024drag}
\bibfield{author}{\bibinfo{person}{Yujun Shi}, \bibinfo{person}{Chuhui Xue}, \bibinfo{person}{Jun~Hao Liew}, \bibinfo{person}{Jiachun Pan}, \bibinfo{person}{Hanshu Yan}, \bibinfo{person}{Wenqing Zhang}, \bibinfo{person}{Vincent~YF Tan}, {and} \bibinfo{person}{Song Bai}.} \bibinfo{year}{2024}\natexlab{}.
\newblock \showarticletitle{Dragdiffusion: Harnessing diffusion models for interactive point-based image editing}. In \bibinfo{booktitle}{\emph{Proceedings of the IEEE/CVF Conference on Computer Vision and Pattern Recognition}}. \bibinfo{pages}{8839--8849}.
\newblock


\bibitem[Wang et~al\mbox{.}(2025)]%
        {wang2024framer}
\bibfield{author}{\bibinfo{person}{Wen Wang}, \bibinfo{person}{Qiuyu Wang}, \bibinfo{person}{Kecheng Zheng}, \bibinfo{person}{Hao Ouyang}, \bibinfo{person}{Zhekai Chen}, \bibinfo{person}{Biao Gong}, \bibinfo{person}{Hao Chen}, \bibinfo{person}{Yujun Shen}, {and} \bibinfo{person}{Chunhua Shen}.} \bibinfo{year}{2025}\natexlab{}.
\newblock \showarticletitle{Framer: Interactive frame interpolation}.
\newblock \bibinfo{journal}{\emph{International Conference on Learning Representations (ICLR)}} (\bibinfo{year}{2025}).
\newblock


\bibitem[Xing et~al\mbox{.}(2024)]%
        {xing2024tooncrafter}
\bibfield{author}{\bibinfo{person}{Jinbo Xing}, \bibinfo{person}{Hanyuan Liu}, \bibinfo{person}{Menghan Xia}, \bibinfo{person}{Yong Zhang}, \bibinfo{person}{Xintao Wang}, \bibinfo{person}{Ying Shan}, {and} \bibinfo{person}{Tien-Tsin Wong}.} \bibinfo{year}{2024}\natexlab{}.
\newblock \showarticletitle{Tooncrafter: Generative cartoon interpolation}.
\newblock \bibinfo{journal}{\emph{ACM Transactions on Graphics (TOG)}} \bibinfo{volume}{43}, \bibinfo{number}{6} (\bibinfo{year}{2024}), \bibinfo{pages}{1--11}.
\newblock


\bibitem[Zhang et~al\mbox{.}(2023)]%
        {zhang2023controlnet}
\bibfield{author}{\bibinfo{person}{Lvmin Zhang}, \bibinfo{person}{Anyi Rao}, {and} \bibinfo{person}{Maneesh Agrawala}.} \bibinfo{year}{2023}\natexlab{}.
\newblock \showarticletitle{Adding conditional control to text-to-image diffusion models}. In \bibinfo{booktitle}{\emph{Proceedings of the IEEE/CVF international conference on computer vision}}. \bibinfo{pages}{3836--3847}.
\newblock


\end{thebibliography}
}

\end{document}